# Phase-matched four-wave mixing in the extreme ultraviolet region

Khoa Anh Tran,[†] Khuong Ba Dinh, Peter Hannaford, and Lap Van Dao

Centre for Quantum and Optical Science, Swinburne University of Technology, Melbourne 3122, Australia

[†]Correspondence author: *anhkhoatran@swin.edu.au*

*Abstract*

We report here a detailed study of the four-wave mixing process in the extreme ultraviolet (XUV) region around 30 nm by using two collinear incommensurate frequency laser pulses. The experimental results reveal evidence of the coherent accumulation of the wave-mixing fields and low-order (third-order and fifth-order) nonlinear response of an argon medium. The dependence of the intensities of the mixing fields on the intensity of a weak control field, on the argon pressure and on the interaction length is analyzed to show that the four-wave mixing fields in this spectral range are generated under the phase-matched condition.

## *I. Introduction*

In the past decade, high-harmonic generation (HHG) has become a promising method to generate coherent radiation in the XUV and soft X-ray region [1,2]. HHG is a highly nonlinear process which up-converts the frequency $\omega_l$ of a fundamental laser into its harmonics $q\omega_l$. The HHG spectrum can range from hundreds of nanometers down to sub-nanometer wavelengths [3-5], and the ultrashort pulse duration [6] (femtosecond down to a few tens of attosecond) of the emitted radiation paves the way for high-resolution time-resolved absorption spectroscopy [7,8], coherent diffractive imaging [9,10], XUV interferometry [11], and autoionization of gases [12,13].

In two-color HHG, the combination of a driving field and a control field can take advantage of the high non-linearity of the medium to create new frequencies. With an appropriate choice of the intensity ratio between an 800 nm pulse and a weak 1,400 nm pulse, intense attosecond pulses can be produced without the need for carrier-envelope phase (CEP) stabilization [14]. As a result, a dense spectrum structure emerges around zero-time delay of the two pulses as a consequence of sum-frequency and difference-frequency processes in the XUV regime [14].

Four-wave mixing (FWM), which is the dominant mechanism responsible for the generation of new frequencies, is a fundamental process in the nonlinear interaction between intense laser light and matter. In centrosymmetric matter, the second-order susceptibility $\chi^{(2)}$ vanishes. Thus, the third-order nonlinear susceptibility $\chi^{(3)}$ is the lowest non-vanishing order responsible for the response of such materials to a laser field.

Stimulated Raman scattering is a popular spectroscopic technique based on such a mechanism [15,16]. A few studies of wave-mixing in the XUV region have been made with two commensurate [17-20] or incommensurate [21-23] frequencies in which a gas jet [17-19], a long gas cell [22,23], or a hollow waveguide [20] is the chosen geometric configuration. New frequencies generated by the nonlinear wave-mixing process are realized to satisfy the conservation laws of momentum and energy [19]. Recently, this process has been claimed to involve a cascaded wave-mixing mechanism in which HHG photons participate in the production of four-wave and six-wave mixing frequencies [23]. With recent progress in this field, the soft X-ray laser can now be shifted to previously inaccessible wavelengths with high temporal and spatial coherence by phase-matched wave-mixing processes.

In this paper, we report studies of the wave-mixing process in the spectral range around 30 nm by using two multiple-cycle and incommensurate frequencies (wavelength 800 nm and 1,400 nm). With a collinear configuration of the two beams, an intense and sharp HHG spectrum is optimized with ~ $2 \times 10^{14}$ W/cm$^2$ 800-nm laser pulses. New mixing-frequencies then emerge when the second weak field at 1,400 nm ($< 5 \times 10^{13}$ W/cm$^2$) is applied to an argon interaction medium. Experimental data shows that a high third-order and fifth-order nonlinear response of the argon gas is the main mechanism responsible for the accumulation of these coherent mixing fields. The dependence of the intensities of the mixing frequencies on the intensity of the control field, on the pressure of the argon gas and on the interaction length is then discussed to confirm that the four-wave mixing fields in this spectral region are generated under the phase-matched condition.

The paper is organized as follows. In Sec. II, we present the experimental configuration used in this study. In Sec. III, we discuss some factors of dispersion influencing the efficient phase-matched HHG and four-wave mixing process. In Sec. IV, we demonstrate proof of the phase-matched four-wave mixing processes in the XUV region by analyzing the influence of the intensity of the control field, the gas pressure, and the interaction length on the generated mixing frequencies. Finally, the conclusions are presented in Sec. V.

*II. Experimental setup*

The experimental layout for this study is shown in detail in Fig. 1. A 1 kHz multi-stage, multi-pass, chirped-pulse amplifier (Odin-II, Quantronix) is used to produce 6.0 mJ, 800 nm, 30 fs laser pulses. The beam is split into two separate beam lines with a beam splitter. The transmitted beam of energy 4.0 mJ is truncated, and used as a driving field for HHG in a 20-cm long cylindrical cell filled with argon gas. The reflected beam of energy 2 mJ is directed into a two-stage optical parametric amplifier (Palitra-C.FS, Quantronix) to create a 0.4 mJ, 1,400 nm, 40 fs control field. HHG is generated by the 800 nm pulse (~ $2 \times 10^{14}$ W/cm$^2$) and the intensity of the 1,400 nm field is kept at a low level ($< 5 \times 10^{13}$ W/cm$^2$) so that it does not create HHG radiation by itself. A half-wave plate is used to control the relative polarization between the two beams. A telescope (two lenses L1, L2)

and lens L3 ($f_3$ = 20 cm) are used both to obtain an optimal flux of the HHG spectrum and

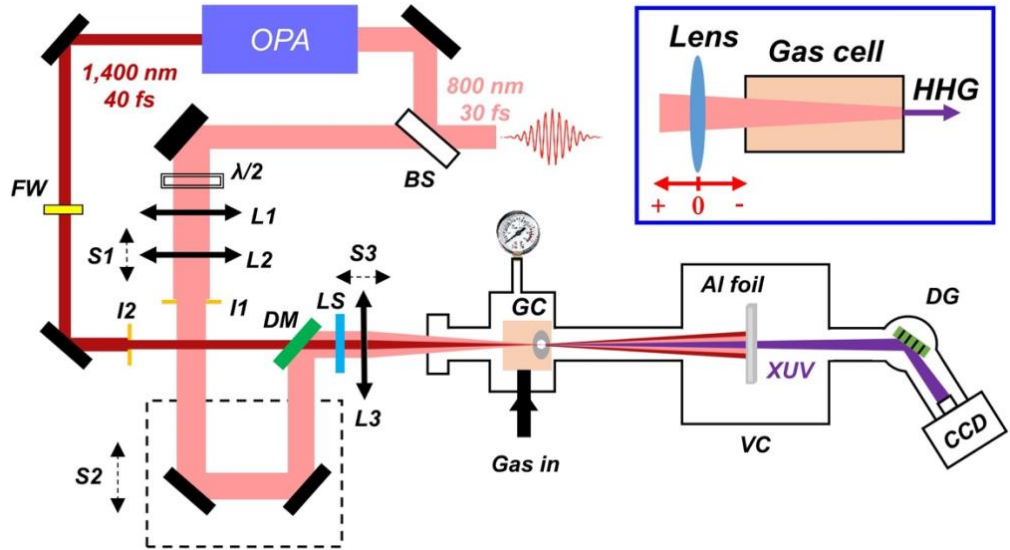

Figure 1. Experimental layout: driving pulse (800 nm, 30 fs, $\omega_1$); control pulse (1,400 nm, 40 fs, $\omega_2$); BS: beam splitter; OPA: optical parametric amplifier; λ/2: half-wave plate; L1, L2, L3: lenses; S1, S2, S3: translational stages; I1, I2: irises; DM: dichroic mirror; LS: laser shutter; FW: filter wheel; GC: gas cell; XUV: emitted radiation; VC: vacuum chamber; DG: diffraction grating; CCD: charge-coupled device. The inset illustrates how the focus position is varied. Positive, zero, and negative values of the focus position imply that the focal point is inside, at the exit plane, and outside of the gas cell, respectively.

to control the focus position of the two pulses. The two laser beams co-propagate when combined with a dichroic mirror DM. The phase-matching condition for the HHG is maintained during the experiment. This condition is obtained by manipulating the diameter of the iris I1 (which truncates the driving laser beam), the relative position of lens L3 to the gas cell and the chirp of the driving laser. The intensity of the control pulse is controlled with a second iris I2 and a filter wheel FW which is installed with five neutral density filters. The pressure of the argon gas in the cell, which varies up to 200 torr, is measured with a pressure gauge Pfeiffer Vacuum D-35614 Asslar. The emitted spectral profile is recorded with a 600 grooves/mm diffraction grating (GIMS-3) and a CCD (XO-PIXIS 1024B, Princeton Instruments). The delay time between the two beams is controlled by a DC motor with 0.1 fs temporal resolution S2 (Newport PM80065). Two lenses, L2 and L3, are installed on two translational stages, S1 and S3, with spatial resolution 24 μm. The stage S3 is a linear actuator (T-LA28A Zaber) which accurately controls the relative focus position of the two laser pulses to the exit plane of the gas cell. A 150 μm pinhole on the exit plane (a ~ 300-μm-thick aluminium plate) of the gas cell, from which emitted radiation generated in the cell propagates toward the CCD camera, is directly drilled by the focused 4-mJ 800-nm beam. A 300-nm-thick aluminium foil (filtering out the residual 800 nm and

1,400 nm beams), a diffraction grating, and a CCD camera are installed in the vacuum chamber. The pressure inside the chamber in the vicinity of the CCD is kept around $10^{-5}$ torr to limit re-absorption of the generated radiation. Vacuum pumps (Pfeiffer, Agilent Technologies) are used to continuously pump gas out of the chamber. The slit, installed at the entrance of the diffraction grating, is opened so that a high visibility of the spectrum is recorded on the CCD. The operating temperature of the CCD sensor (cooled down by air) is set to about $10^0 C$ or lower to reduce background noise in the spectrum. The CCD is connected to and controlled by a computer so that the exposure time of the CCD chip can be easily changed to properly meet the requirements of the experiment. The internal signal from the CCD simultaneously triggers the operation of the two translation stages S2 and S3 and the laser shutter. The diameter of the two beams at the focus is about 150 μm.

In this report, we discuss the experimental data obtained with parallel polarization between the two beams. A negative delay time implies that the 1,400 nm pulse precedes the 800 nm pulse. Zero time delay is therefore assumed to be when the two pulses temporally overlap. Intuitively, the photon flux of all odd harmonic orders produced by the driving laser pulse is a minimum at zero-delay.

### III. Theoretical background

#### 1. Phase-matched generation of high-order harmonics

For the generation of high-order harmonics, a high intensity driving laser field is required to produce some ionization in the interaction medium. On a macroscopic scale, usable HHG radiation comprises a bunch of single photons of the same frequency. All of these single-frequency components must add up constructively to accumulate a coherent laser-like short-wavelength source. Due to wave-vector mismatch $\Delta k_q$ between the fundamental driving laser ($\lambda_l$) and the harmonic radiation ($\lambda_q$), such a source can only be built-up under a phase-matched condition when traversing a nonlinear interaction medium. The total phase mismatch for the $q^{th}$ harmonic order is

$$\Delta k_q = k_q - qk_l = \Delta k_{neutral} + \Delta k_{plasma} + \Delta k_{geom} + \Delta k_{dipole}. \tag{1}$$

The first term is due to the high density of the neutral gas making up the interaction medium $\Delta k_{neutral}$; the second term arises from free electrons released from the atoms in the HHG process $\Delta k_{plasma}$; the third term represents the phase-shift incurred by the geometric configuration of the experiment $\Delta k_{geom}$; and the final term is the intrinsic, intensity-dependent dipole phase of the $q^{th}$ harmonic order $\Delta k_{dipole}$.

The wave-vector $k(\lambda) = n(\lambda)\omega/c$ depends on the refractive index $n(\lambda)$, where c is speed of light in vacuum. The neutral dispersion

$$\Delta k_{neutral} = n(\lambda_q)\omega_q/c - qn(\lambda_l)\omega_l/c = q\omega_l/c[n(\lambda_q) - n(\lambda_l)] \tag{2}$$

where $\omega_q = q\omega_l$. This contribution is negative because the refractive index of the harmonic radiation $n(\lambda_q)$ in the XUV region is always smaller than that of the near-infrared driving field $n(\lambda_l)$.

High-harmonic generation occurs in a highly nonlinear medium. A large free electron density is produced but most of the electrons do not return to the parent ions. The propagation time of these charge particles out of the focal region is much longer (nanoseconds) than the duration of laser pulse (femtoseconds). The plasma frequency is given by $\omega_p = e\sqrt{N_e/(\epsilon_0 m_e)}$, where $\epsilon_0$ is the dielectric constant, $m_e$ the electron mass, $e$ the charge of the electron, and $N_e$ the density of the free electrons. This resonance leads to a polarizability of the plasma which in turn produces the refractive index [24]

$$n_{plasma}(\omega) = \sqrt{1 - \frac{\omega_p^2}{\omega^2}} = \sqrt{1 - \frac{N_e}{N_c(\omega)}} \qquad (3)$$

where $N_c(\omega) = (\epsilon_0 m_e \omega^2)/e^2$ is the critical plasma density at which the plasma medium completely absorbs all the electromagnetic waves of frequency $\omega$. With a driving field of 800 nm, the free-electron density generated is much lower than the critical density ($\sim 2 \times 10^{21}$ cm$^{-3}$). Therefore, we can use approximation

$$n_{plasma}(\omega) \approx 1 - \frac{1}{2}\frac{\omega_p^2}{\omega^2} \qquad (4)$$

The plasma contributes to the wave-vector mismatch

$$\Delta k_{plasma} = \frac{q\omega_l}{c}[n_{plasma}(\lambda_q) - n_{plasma}(\lambda_l)] = \frac{\omega_p^2}{2qc\omega_l}(q^2 - 1) \qquad (5)$$

The free-electron plasma makes a positive contribution to the total phase-mismatch for harmonic orders $q \gg 1$. In our studies, the Rayleigh length of the driving laser (5 to 20 mm) is much longer than the interaction length (2 to 4 mm). Therefore, the phase-matching condition for efficient HHG can be achieved when the neutral dispersion counterbalances the plasma dispersion without considering dispersion contributed by the Gouy phase shift and the intrinsic harmonic dipole phase [25]:

$$\Delta k_q = q\omega_l/c[n(\lambda_q) - n(\lambda_l)] + \omega_q^2/(2qc\omega_l)[q^2-1] \qquad (6)$$

The intensity of a phase-matched harmonic order in a dispersive and absorbing medium is then estimated to be [26-28]

$$I_q \approx N^2 L^2 |d^{NL}_q|^2 \exp(-\alpha_q L/2)[\sin^2(\Delta k_q L/2) + \sinh^2(\alpha_q L/4)]/[(\Delta k_q L/2)^2 + (\alpha_q L/4)^2] \qquad (7)$$

In Eq. (7), re-absorption of the harmonic signal by the gas medium is taken into account,

$d^{NL}{}_q$ is the amplitude of the nonlinear atomic dipole moment, and $α_q$ is the XUV absorption coefficient of the generating medium. The harmonic intensity $I_q$ reduces to the familiar $sinc^2(\Delta k_q L/2)$-dependence when the absorption coefficient $α_q$ is small. We note that the magnitude $I_q$ of a phase-matched $q^{th}$-order harmonic varies quadratically with the gas pressure and the interaction length. Therefore, an experimental investigation of how $I_q$ varies with $p$ and $L$ can quantitatively provide proof of a phase-matched HHG process.

## 2. Phase-matched four-wave mixing

The generation of mixing waves is a coherent process over the interaction length $L$. Therefore, the intensity of a four-wave mixing field [15]

$$I_4 \sim I_1 I_2 I_3 |χ^{(3)}|^2 N^2 L^2 sinc^2(\Delta k L/2), \tag{8}$$

where $I_1$, $I_2$, $I_3$ are the intensities of the driving, control and HHG fields, $N$ is the gas density, and $\Delta k$ is the phase mismatch for the generation of the four-wave mixing fields

$$\Delta k = k_4 - k_3 \pm (k_1 - k_2). \tag{9}$$

The strength of these mixing fields is proportional to the intensities of all constituent fields, the square of the atomic density $N^2$, the square of the interaction length $L^2$, and the function of phase mismatch $\Delta k L$. If the wave-mixing process is induced with constant driving fields and a small phase mismatch, the development trend of mixing frequencies increases quadratically with $N$ and $L$. In such a case, the four-wave mixing process is also phase-matched.

## IV. Results and discussion

### 1. Phase-matched generation of high-order harmonics

In this section, some aspects of the HHG spectrum generated with an 800-nm driving pulse are discussed when we vary either the interaction length or the gas pressure. The radius of the focal point can be varied (by the iris I1) from 45 μm to 100 μm over which the Rayleigh length varies from 5 mm to 20 mm and the effective focused intensity is $\sim 2 \times 10^{14}$ W/cm$^2$. The HHG is driven by the 800-nm laser pulse (carrier frequency $ω_1$). A typical HHG spectrum is shown in the - 150 fs plot of Fig. 2. The large delay time - 150 fs between the two pulses or absence of the second beam is chosen because the effect of the control field (carrier frequency $ω_2$) on the HHG process is negligible. Therefore, the spectrum at - 150 fs comprises only odd-harmonic orders, i.e., 21 to 33 (H21 to H33). A few high-flux, narrow-bandwidth odd harmonics are produced under the current experimental conditions. The sharpness and spectral width of the harmonics are found unchanged with the time delay between the two pulses, e.g., at 0 fs and - 150 fs. The inset of Fig. 2 shows beam profiles with good spatial coherence for the three most intense harmonics, i.e., H25 (black solid line), H27 (red dotted line) and H29 (green dashed line).

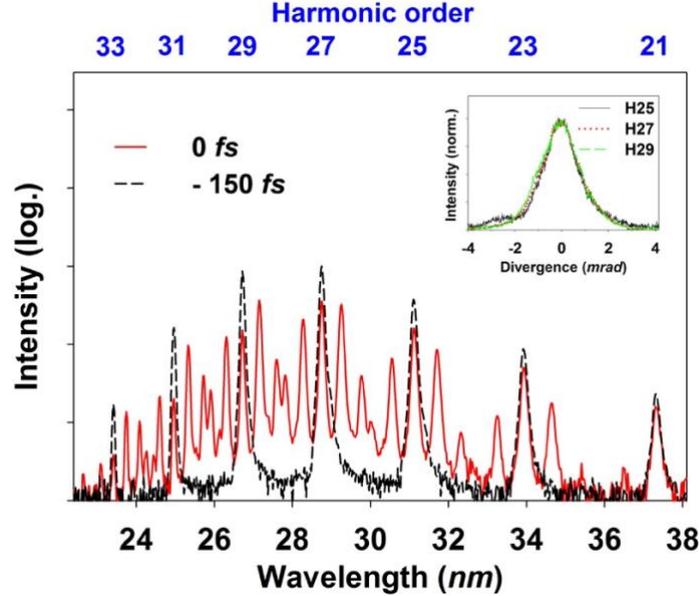

Figure 2. Illustration of spectra obtained with 60 torr of argon gas at 0 fs (red solid line) and - 150 fs (black dashed line). The spectrum at - 150 fs only comprises odd harmonic orders (i.e., H21 to H33). However, the spectrum at 0 fs exhibits both odd harmonic orders and four additional mixing peaks on either side of each harmonic. The inset shows beam profiles with good spatial coherence of three harmonic orders, i.e., H25 (black solid line), H27 (red dotted line), and H29 (green dashed line) taken at delay time - 150 fs.

An intense and narrow HHG spectrum is also observed when the focus position of the driving laser pulse is moved relative to the exit of the gas cell filled with argon gas at 60 torr, Fig. 3(a), and at 120 torr, Fig. 3(b). In each case, we define the position x = 0 where the focus position is at the exit of the gas cell. The intensity of the driving field is then optimized for a maximum output intensity of all small-bandwidth and Gaussian-profile high-order harmonics. With such a yield, the position of the focal point is varied for a change of the interaction length while all other parameters are kept constant. Positive and negative values of the focus position consequently imply that the focal point is inside and outside of the gas cell, respectively (see inset of Fig. 1). As can be seen in Figs. 3(a) and 3(b), a few bright and sharp harmonic orders are generated. However, harmonic orders up to H35 are generated with a pressure of 120 torr rather than H31 with 60 torr. In Figs. 3(c) and 3(d), the two strongest orders, i.e., H25 (black solid circles) and H27 (red triangles), H27 (green diamonds) and H29 (blue squares), which are extracted from Figs. 3(a), and (b), are illustrated. The development trend of these harmonics is well-fitted with Eq. (7), where the absorption and phase mismatch, which is dependent on the interaction length, are considered. When the focus position x is < 3 mm (at 60 torr) and < 1 mm (at 120 torr), the intensity of the harmonics increases quadratically, Fig. 3. Hence, the phase mismatch is very small in this interaction length at these two pressures. In this measurement, the effective interaction length is inferred from the displacement of the focus position over which the phase matching condition for HHG is satisfied. Thus, the interaction lengths of the harmonic orders at 60 torr and 120 torr are ~ 4.5 mm and 2 mm, respectively. As a result, the ratio of increasing slopes of HHG intensities for the two pressures is ~ 2.25, which is approximately equal to the ratio of the gas pressures (120/60 = 2). We note that the interaction length for all harmonic orders is almost constant for a specific gas density, and the intensity of the harmonics below H27 (> 30 nm), e.g., H21 and H23 (Fig. 3(a)), H23 and H25 (Fig. 3(b)), is seen to quickly decrease when the position of the laser focus moves deeper into the gas cell. This is due to a large phase mismatch and the influence of the absorption of the gas medium. When x > 4 mm, a dominant re-absorption in the interaction medium attenuates the signal of the harmonics by an exponential decrease of the interaction length.

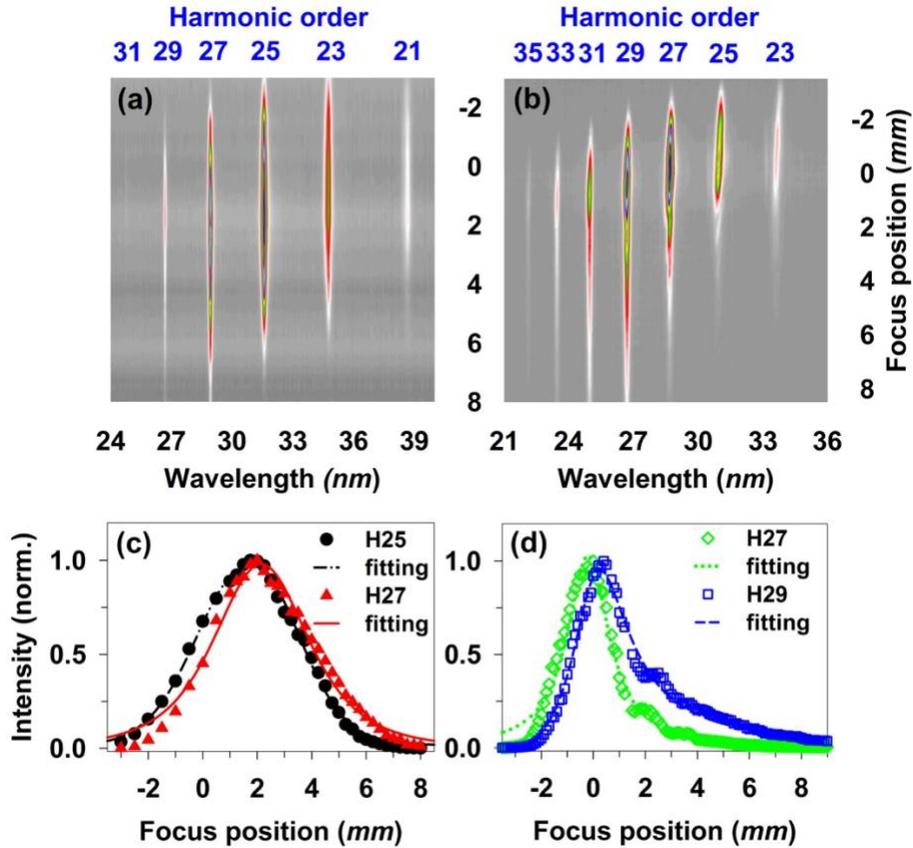

Figure 3. HHG spectra taken at delay time - 150 fs with argon at 60 torr (a) and 120 torr (b), when the focus position is scanned from outside (negative value) to inside of the gas cell (positive value). Intensities of the two strongest harmonics at 60 torr (H25 - black solid circles, H27 - red triangles), and 120 torr (H27 - green diamonds, H29 - blue squares), which are extracted from plots (a) and (b), are shown in (c) and (d), respectively. The lines accompanying the scattered plots are fitting curves of the corresponding harmonics, using Eq. (7).

The phase matching of HHG is further examined by considering the variation of the harmonic spectrum versus argon gas pressure. The experimental conditions are separately optimized with a pressure of 60 torr, Fig. 4(a), and 120 torr, Fig. 4(b), so that a strong and sharp spectrum of all available harmonics is produced. The gas pressure in these two cases is then tuned between 20 and 100 torr, Fig. 4(a), and between 55 and 200 torr, Fig. 4(b), while other experimental parameters are fixed. The intensity of the harmonic grows between the lowest gas density and up to the chosen optimal pressure, then decreases as the pressure keeps increasing. However, a sharp and narrow spectral width of all harmonic orders is maintained as the gas pressure is varied. We also note that the development of the intensity of the harmonics around H27 (~ 29 nm), i.e., when the pressure increases up to about 60 torr (120 torr), the generation of the H27, H29 (H29, H31) orders is favorable but the intensities of the H21, H23 (H23, H25) orders are strongly diminished. This is due to the strong absorption of argon gas in the spectral range > 30 nm, Fig. 4(a) (Fig. 4(b)).

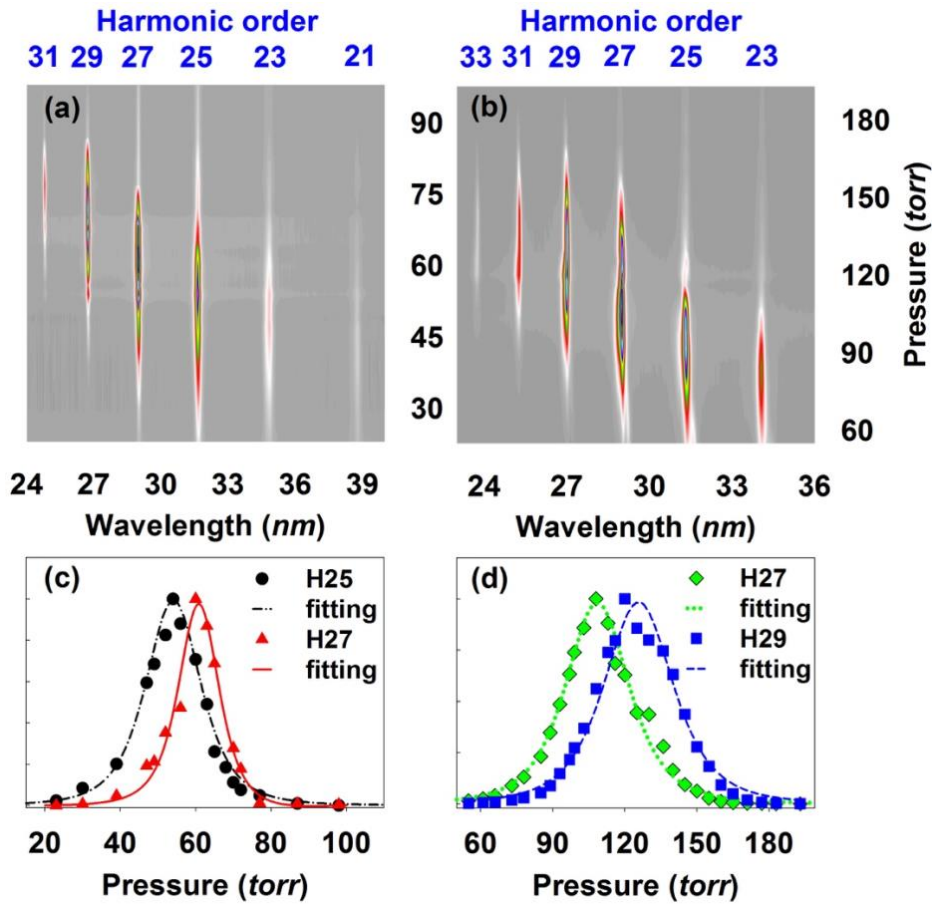

Figure 4. Dependence of the intensity of the harmonic radiation on the argon gas pressure. The experiments are optimized at 60 torr (a) and 120 torr (b). The intensities of the two strongest harmonics at 60 torr (H25 - black solid circles, H27 - red triangles), and 120 torr (H27 - green diamonds, H29 - blue squares), which are extracted from plots (a) and (b) are shown in (c) and (d), respectively. Data is taken at a delay time of - 150 fs. The lines accompanying the scattered plots are fitting curves of the corresponding harmonics, using Eq. (7).

The two intense harmonic orders, H25 (black solid circles) and H27 (red triangles), Fig. 4(c), and H27 (green diamonds) and H29 (blue squares), Fig. 4(d), which are extracted from the two plots shown in Figs. 4(a) and (b), respectively, are plotted on the same graph for a discussion of the phase-matching condition in the HHG production. The development trend of these intensity profiles fits closely with the model given by Eq. (7). However, there is a significant shift between intensity profiles of the two chosen harmonics. This is caused by the effect of absorption of the gas medium on the different harmonic orders. In Fig. 4(c), as the pressure $p < 54$ torr for H25, and $p < 60$ torr for H27, the strength of the H25 and H27 orders increases quadratically with pressure. Similarly, the signal of H27 and H29 also increases with the square of the pressure over the range $p < 110$ torr and $p < 120$ torr, respectively, Fig. 4(d). This is evidence of phase-matched HHG in a strongly absorptive medium. As the pressure continues to increase, i.e., $p > 54$ torr for H25 and $p > 60$ torr for H27 (Fig. 4(c)), $p > 110$ torr for H27 and $p > 120$ torr for H29 (Fig. 4(d)), the efficiency of the harmonic generation surpasses the absorption limit at which the intensity exponentially decays with pressure.

In conclusion, sharp and narrow bandwidth HHG is produced in our investigation of interaction length-dependence and pressure-dependence. The corresponding interaction lengths, i.e., 2 mm, and 4.5 mm, are both smaller than the estimated Rayleigh length (from 5 mm o 20 mm), and there is a significant $L^2$-dependence, and $p^2$-dependence of the HHG intensities when the experimental conditions are optimized at 60 and 120 torr. Thus, HHG is generated with phase-matched condition.

## *2. Phase-matched four-wave mixing*

In Fig. 2, the central wavelengths of all harmonics are labelled with corresponding orders, i.e., H21 to H33. In addition to the main harmonics, four extra frequencies are recorded on either side of each odd harmonic at zero time delay. Moreover, the spectra of these new wavelengths are as sharp as those of the main harmonics. The sharpness of both the main harmonics and these new frequencies is also found and preserved when the interaction length (Figs. 5(a), (b)) or gas pressure (Figs. 5(c), (d)) is changed. This is a reliable signature of wave-mixing processes in the XUV region. For simplicity, the carrier frequency ω of a laser field also denotes its corresponding photon energy. Therefore, the photon energy difference of the driving field and the control field, $\Delta\omega = \omega_1 - \omega_2$, is roughly 0.67 eV. On each side of one harmonic, the energy gap between it and the first neighboring peak (first mixing-order) and between the first and a further peak (second mixing-order) is also estimated to be ~ 0.67 eV (plot 0 fs of Fig. 2, and Fig. 5). Given that the energy of the $q^{th}$-order harmonic is $q\omega_1$, the corresponding energy of the mixing waves $\omega_{mix}$ is then approximately $q\omega_1 \pm m\Delta\omega$ ($m = 1, 2$), where the "plus" and "minus" of $m\Delta\omega$ represent the two closest peaks to the left (shorter wavelength) and to the right (longer wavelength) of the $q^{th}$ order. The generation of these new frequencies is claimed to be the sum-frequency mixing (SFM) and difference-frequency mixing (DFM) processes in the XUV region, respectively [17].

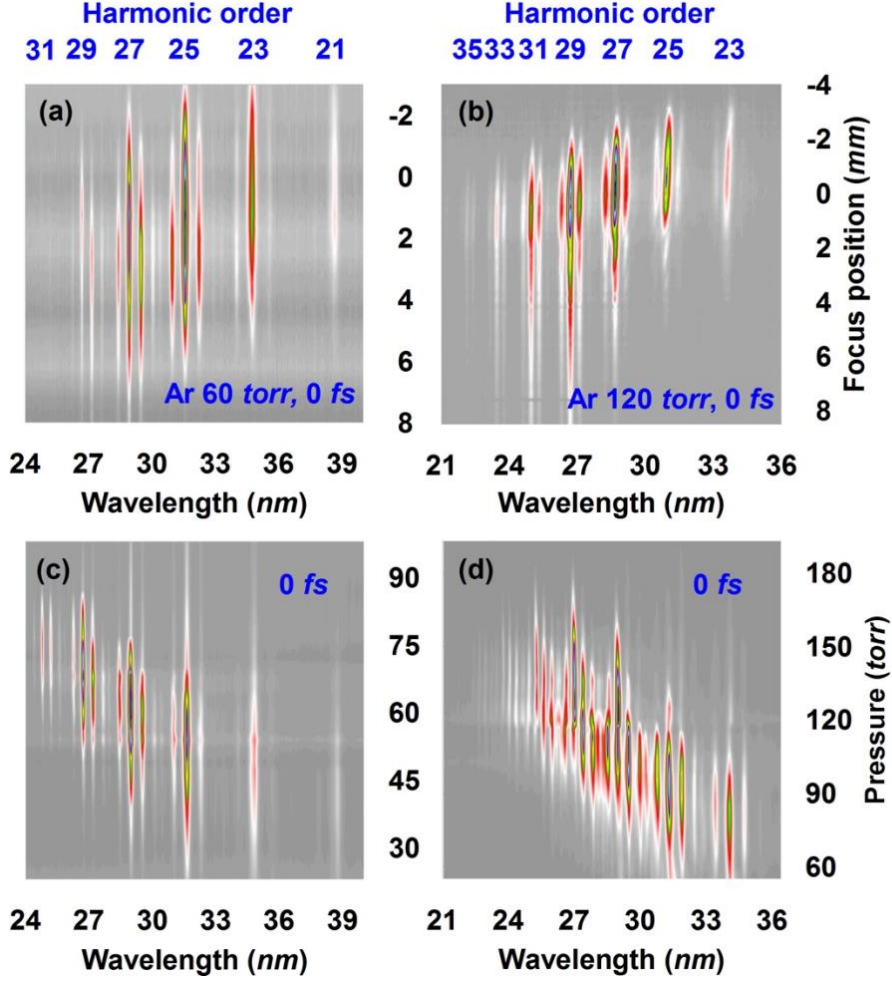

Figure 5. Illustration of sharp and constant spectral bandwidth of both harmonics and wave-mixing frequencies generated at a time delay 0 fs with argon gas at a pressure of 60 torr (a) and 120 torr (b) as a function of focus position. In (c) and (d), the experimental condition is optimized with argon at 60 torr and 120 torr before the pressure in the cell is varied. These results are taken from the same data sets as shown in Figs. 3 and 4, where the spectra recorded at - 150 fs time delay are discussed.

Let $\omega_3 \equiv q\omega_1$, $\omega_4 \equiv \omega_{mix}$, $k_3 \equiv qk_1$, $k_4 \equiv k_{mix}$ be the carrier frequencies and the wave-vectors of the $q^{th}$ harmonic order and the mixing field, respectively. $\omega_3$ is generated under the phase-matching condition as discussed in Section IV. 1. The experimental data reveals that the frequency of the mixing wave

$$\omega_4 \simeq \omega_3 \pm m(\omega_1 - \omega_2). \tag{10}$$

The spatial and temporal coherence of the driving pulse is transferred to the HHG radiation [29]. The phase of the harmonics remains unchanged over the interaction length because the neutral and plasma dispersion is small for an XUV pulse, and $(k_1 - k_2) \ll k_3, k_4$. Thus, the total wavevector mismatch for the generation of the mixing field is

$$\Delta k = k_4 - k_3 \pm m(k_1 - k_2) \approx 0. \qquad (11)$$

Equations (10), and (11) therefore satisfy the known energy and momentum conservation laws [19,22].

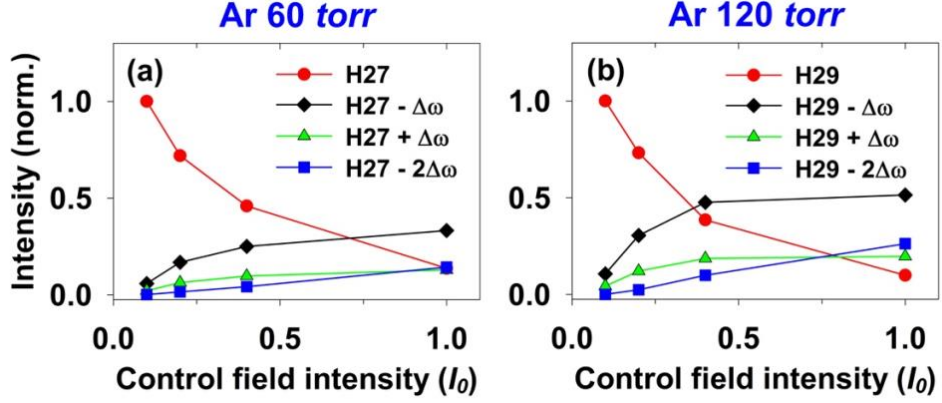

Figure 6. Intensity of the strongest harmonic order and corresponding mixing frequencies versus relative intensity of the control field $I_2$ with argon gas at 60 torr (a) and 120 torr (b).

The SFM and DFM processes occur in the presence of the second field $\omega_2$; therefore the intensity of the mixing field should obey Eq. (8). The variation of the intensity of the strongest harmonic order and the mixing frequencies versus the relative intensity of the control field $I_2$ at 60 torr and 120 torr is then examined and is shown in Fig. 6(a) (H27 and H27 ± $m\Delta\omega$), and Fig. 6(b) (H29 and H29 ± $m\Delta\omega$). The experimental data in these two plots is extracted at zero time-delay between the $\omega_1$ and $\omega_2$ fields. Also, the highest intensity of the control field $I_0$ is controlled below the threshold at which the signal of the first-order mixing waves start to saturate. We observe a similar trend of the intensities between H27 and H29 and between H27 ± $m\Delta\omega$ and H29 ± $m\Delta\omega$, i.e., a depletion of the main harmonic order while the intensity of all mixing fields increases. When $I_2$ is low ($\leq 0.2I_0$), the increasing intensity of the mixing fields and the decreasing intensity of the odd harmonics scale monotonically with $I_2$ (as expected from Eq. (8)). However, for moderate intensity of the control field (from $0.2I_0$ to $0.4I_0$), the second-order mixing field continues to increase linearly with $I_2$ but the increment of the first-order intensity and the decline of the HHG intensity is no longer linear. At these intensities of the control field the intensity of the harmonics is low; therefore there is a decrease of the intensity of $I_3$ in Eq. (8). Additionally, the higher-order nonlinear response of the interaction medium also needs to be taken into account. A similar development of the intensities of other harmonics and their corresponding mixing fields is also observed in the study.

In summary, with the power scaling of the mixing fields, the depletion of the original odd harmonics, the correlation between the intensities of the original and new fields, Eq. (8), the allowed photon combinations contributing to the new fields (see Eq. (10)), the nonlinear optical wave-mixing processes in the XUV region involving a cubic- (m = 1) and a fifth-order (m = 2) nonlinear susceptibility are attributed to the above-mentioned observation. We note that it is not possible to study the power scaling of the second-order mixing field because this requires a high-intensity control field. When a high-intensity control field is applied, other nonlinear processes need to be considered. However, the decay of the second-order mixing signal for a long positive delay may provide indirect evidence of a fifth-order nonlinearity. Therefore, Eq. (8) can be modified as follows for dealing with nonlinear processes involving a fifth-order nonlinear susceptibility $\chi^{(5)}$, $I_5 \sim I_1^2 I_2^2 I_3 |\chi^{(5)}|^2 N^2 L^2 sinc^2(\Delta k L/2)$, where $\Delta k$ is mentioned in Eq. (11).

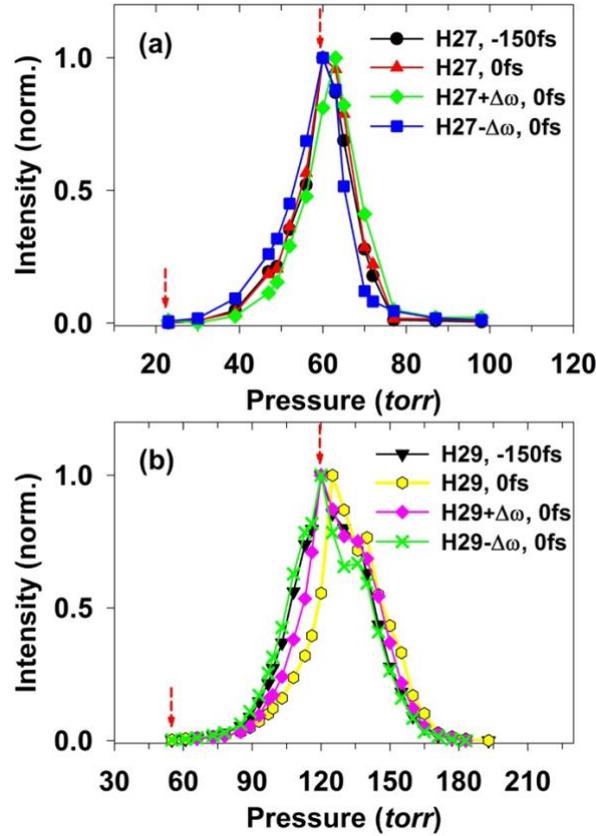

Figure 7. Dependence of the intensity of the harmonic orders and mixing fields on argon gas pressure. The experimental conditions are optimized around pressures of 60 torr (a), 120 torr (b). The two red dashed arrows in each figure indicate the region where the mixing fields are generated under the phase-matched condition.

The influence of the phase-mismatch on the generation of the mixing fields is investigated by considering the dependence of the output intensities of these fields on the pressure of the argon gas. The experimental condition for the data shown in Fig. 7 is the same as that shown in Fig. 4 and Figs. 5(c) and (d). Here, we compare the signal of the main harmonic at - 150 fs, the main harmonic and the mixing waves at 0 fs in order to see whether or not the phase-matched condition for the wave-mixing processes is fulfilled. The signal of all harmonics (H27 optimized at 60 torr and H29 optimized at 120 torr) is generated with the phase-matched condition (Section IV. 1). For $p < 60$ torr, the development trend of the H27 and H27 $\pm \Delta\omega$ at 0 fs coincides with that of H27 at - 150 fs, Fig. 7(a). This indicates that H27 $\pm \Delta\omega$ are generated with the phase-matched condition. When the mixing wave is produced with the experimental conditions optimized around 120 torr, the signal of the H29 $\pm \Delta\omega$ at 0 fs ($p < 120$ torr) also follows that of H29 at - 150 fs. Therefore, the signal of H29 $\pm \Delta\omega$ is produced with the phase-matched condition. However, the phase-matched condition for H29 at 0 fs only holds for $p < 100$ torr, Fig. 7(b). At pressures higher than the optimized value, the strength of all harmonic orders and mixing waves is dominated by an exponential decay of the gas pressure due to re-absorption in the gas medium.

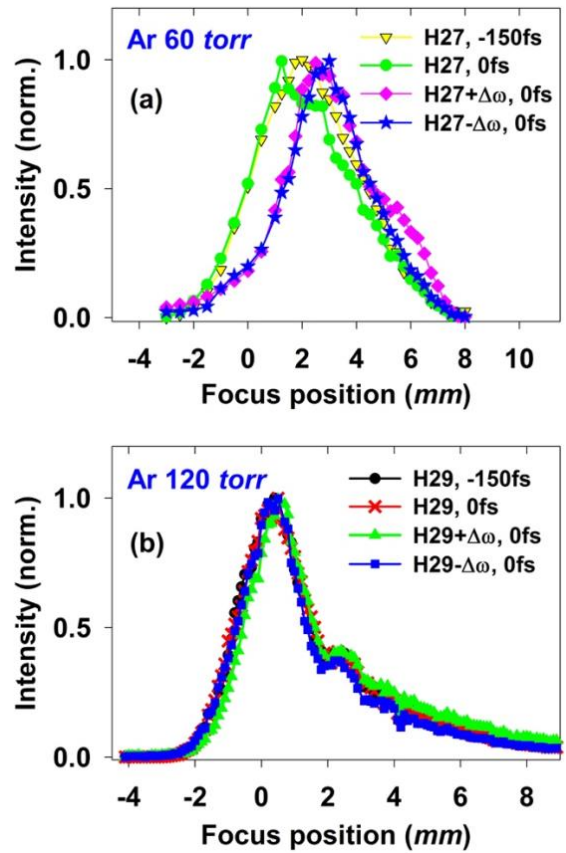

Figure 8. Interaction-length dependence of the intensity of the harmonic orders and mixing fields with argon pressure of 60 torr (a) and 120 torr (b).

The effect of phase mismatch on the intensities of the mixing fields is also studied by varying the interaction length $L$ (scanning stage S3) at two different argon gas pressures 60 torr, Fig. 8(a), and 120 torr, Fig. 8(b). The signal of H27 at 60 torr and H29 at 120 torr, and their corresponding first-order mixing waves, versus focus position is chosen for this discussion. The experimental data at these frequencies are extracted from the same data set that we already demonstrated in Fig. 3 and Figs. 5(a), (b). We include the intensity profiles taken at delay - 150 fs of the two harmonics, i.e., H27 and H29 in Fig. 8, so that the effect of phase mismatch on the mixing-wave signal can be better viewed as the interaction length is changed. These odd harmonics are generated with the phase-matching condition, Section IV. 1. In Fig. 8(a), the intensity of the H27 and H27 $\pm \Delta\omega$ at 0 fs only follows that of the H27 at - 150 fs when $x < $ - 2 mm. When $x > $ - 2 mm, the signal of these mixing waves lags that of the main harmonics. In other words, the interaction length of the mixing frequencies is shorter than that of the original odd harmonics. This could be additional evidence for the participation of the harmonic photons in the production of new frequencies - a cascaded wave-mixing process [23]. However, this phenomenon is not resolvable at 120 torr, Fig. 8(b), when the trend of the mixing waves is almost inseparable from that of the harmonics. This is due to the short interaction length, i.e., about 2 mm, of all generated frequencies. Hence, the wave-mixing frequencies generated at 60 torr as $x < $ - 2 mm and at 120 torr argon are also generated with the phase-matching condition when their intensity profiles closely follow those of the main harmonic orders.

## *V. Conclusions*

We have discussed the phase-matched four-wave mixing processes in the extreme ultraviolet with a collinear two-color HHG configuration. The experimental results reveal clear evidence of a coherent accumulation of the new frequencies and a high third-order nonlinear response of an argon medium. The experimental scheme is compact, and the wave mixing process in the XUV region can be induced with a weak control field. Therefore, the outcome of this study will be applicable for future research that requires efficient production of coherent XUV and soft X-ray sources for high energy ultrafast nonlinear spectroscopy.

### *Acknowledgements*

Financial support was provided by the ARC Discovery Project Scheme (Grant ID DP170104257).